# Microscopic observation of carrier-transport dynamics in quantum-structure solar cells using a time-of-flight technique

Kasidit Toprasertpong[1], Naofumi Kasamatsu[3], Hiromasa Fujii[1], Tomoyuki Kada[3], Shigeo Asahi[3], Yunpeng Wang[2], Kentaroh Watanabe[2], Masakazu Sugiyama[1], Takashi Kita[3], and Yoshiaki Nakano[1]

[1]School of Engineering, the University of Tokyo, Bunkyo-ku, Tokyo 113-0032, Japan
[2]Research Center for Advanced Science and Technology, the University of Tokyo, Meguro-ku, Tokyo 153-8904, Japan
[3]Graduate School of Engineering, Kobe University, Nada-ku, Kobe 657-8501, Japan

In this study, we propose a carrier time-of-flight technique to evaluate the carrier transport time across a quantum structure in an active region of solar cells. By observing the time-resolved photoluminescence signal with a quantum-well probe inserted under the quantum structure at forward bias, the carrier transport time can be efficiently determined at room temperature. The averaged drift velocity shows linear dependence on the internal field, allowing us to estimate the quantum structure as a quasi-bulk material with low effective mobility containing the information of carrier dynamics. We show that this direct and real-time observation is more sensitive to carrier transport than other conventional techniques, providing better insights into microscopic carrier transport dynamics to overcome a device design difficulty.

Quantum structures have been extensively studied as candidate materials for next-generation solar cells[1], such as lattice-matched multijunction[2,3], intermediate-band[4,5], hot-carrier[6,7], and multiple-exciton-generation solar cells[8,9]. However, the obstruction of carriers by the potential barriers degrades the collection efficiency of the photoexcited carriers, and hence, the cell performance[10-12]. In many applications such as lasers[13-15], detectors[16] to amplifiers[17], a number of researchers have focused on the study of the carrier dynamics in quantum structures in order to fully utilize their potential for device performance enhancement. Particularly in solar cells, the device performance is sensitive to the potential barriers at the hetero-interfaces. This is because the photoexcited charge carriers have to be separated and collected as photocurrent, while the electric field has to be weakened to obtain a photovoltage.

So far, open-circuit voltage ($V_{oc}$) and fill factor have been mainly used to evaluate the influence of potential barriers on carrier collection[11,18,19]. The external quantum efficiency (EQE) measurement at different bias voltages is an alternative method proposed to calculate the carrier collection efficiency[12,20,21]. Although these techniques can be easily used to assess the impact on device performance, they do not provide a detailed microscopic understanding of nanostructure physics and are limited to trial-and-error studies. Time-resolved photoluminescence (PL) is one main method to study photoexcited carriers microscopically from the transient response[13,21-23]. However, this technique cannot be used to study carrier-cascade dynamics in periodic structures, namely superlattices, because carriers at different positions generate similar signals and hence the spatial information is difficult to resolve[24]. A more sophisticated technique is required to measure further important microscopic parameters and to reveal the transport dynamics inside quantum-structure solar cells, which will eventually lead to the design optimization towards the enhancement of photocarrier collection.

In this paper, we propose a *carrier time-of-flight* (TOF) technique to evaluate the carrier transport time across a given quantum structure by placing a single-quantum-well probe under the target quantum structure. The schematic is shown in Fig. 1(a). Only the carriers in the vicinity of the top surface [marked as (I) in Fig. 1(a)] are excited by pulse illumination. Charge carriers are separated and driven by the electric field in the active region (II) before traveling across the quantum structure. Some carriers that reach the probe well radiatively recombine and generate PL signal (III). By designing the probe well such that its emission is spectrally separated from the quantum structure, the time taken by the carriers to reach the probe can be observed. On comparing the detected time with that of a *reference* sample whose quantum structure is replaced by the bulk material with the probe well inserted at the same depth [Fig. 1(c)], it is possible to cancel out the time delay caused by the measurement system and carrier dynamics inside the probe, which has not been sufficiently considered in most of other time-of-flight experiments. Hence, the delay originating only from the carrier transport—carrier TOF—can be determined.

The optical TOF techniques proposed in earlier works have tried to measure carrier mobility[25–31] by observing the ambipolar diffusion transport in the flat band region, and most of them were conducted at low temperatures. These techniques cannot be applied to photovoltaic devices which include quantum structures in their thick intrinsic regions. In such devices, the carriers at temperatures equal to or greater than room temperature are driven by electric field and drift transport becomes dominant. In unipolar transport, the lack of the opposite charge carriers in the probe well prevents carriers from recombining, decreasing the probe signal intensity. Additionally, the radiative recombination becomes ineffective at high temperatures because of the increasing non-radiative recombination component.

The ineffective recombination is solved by applying forward bias during the measurement. Under forward bias, both the charge carriers, electrons and holes, are steadily supplied to the probe well in the form of injection current. This was tested on an InGaAs/GaAsP multiple-quantum-well (MQW) p-i-n cell grown by metal-organic vapor phase epitaxy[32] as shown in Fig. 1(b). Fig. 2(a) shows the continuous-wave PL measurement result at 298 K using a 405-nm-wavelength laser with an absorption depth less than 20 nm to selectively excite the sample surface. The PL intensity from the probe well rises by three orders of magnitude on increasing the bias voltage from 0.5 to 0.8 V, similar to the injection current which also increases by the same magnitude. On the other hand, the PL intensity from GaAs and MQW increases by less than two times. This is an evidence for the hole concentration bottleneck in the probe well because the photoexcited holes at the p-doped top surface are blocked by the electric field and MQW, recombining in GaAs and MQW but not in the probe well. Furthermore, applying forward bias provides more accurate information of carrier transport near the solar cell operating voltage.

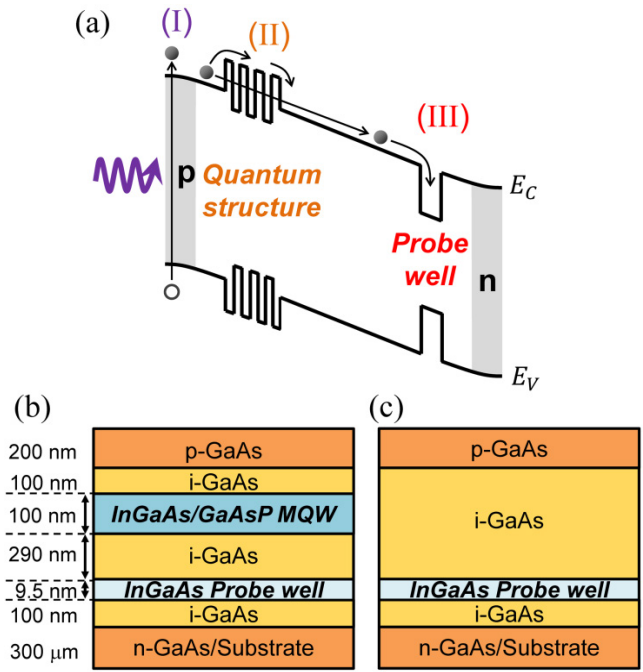
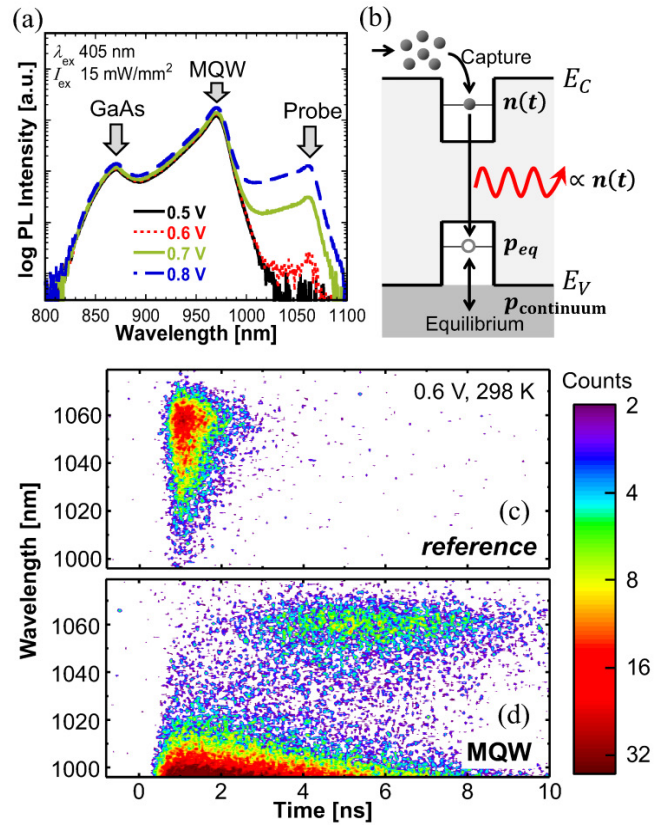

FIG. 1. (a) Schematic of the time-of-flight measurement. Charge carriers are selectively photoexcited near the top surface by a pulse excitation, driven through the quantum structure by the electric field, and detected by a probe well. (b) Structure of the MQW sample used in this work. The device is a p-i-n structure containing an 8-well $In_{0.21}Ga_{0.79}As$ (5.5 nm)/$GaAs_{0.61}P_{0.39}$ (6.2 nm) MQW and a 9.5-nm thick $In_{0.25}Ga_{0.75}As$ probe well in the i-region. The number of GaAsP barrier layers in the MQW was 9, more than InGaAs well layers by 1 for symmetry. In addition, a 25 nm-thick InGaP window layer was grown on the p-type GaAs to passivate the top surface. (c) Structure of the *reference*. MQW was replaced by GaAs bulk with the same thickness.

FIG. 2. (a) PL spectra from the MQW sample after 405-nm-wavelength continuous-wave excitation at various applied bias voltages. The PL peak wavelengths of GaAs, MQW, and the probe well are 871 nm, 969 nm, and 1060 nm, respectively. (b) Carrier dynamics inside the probe well after a pulse excitation. PL after pulse excitation from (c) the *reference*, which is the sample without MQW, and from (d) the MQW sample at 0.6 V measured by a streak camera.

Fig. 2(b) shows the dynamics in the probe well during the time-resolved measurement. After pulse excitation, electrons travel towards the probe well and, subsequently, recombine radiatively. The temporal fluctuation in the hole concentration in the probe on the arrival of photoexcited electrons is negligible owing to the thermal equilibrium with the high-concentration holes in the continuum valence band. This approximation is valid since the thermalization of carrier distribution usually occurs in hundreds of femtoseconds or faster[33,34], whereas the recombination in quantum structures generally occurs on a timescale in the order of nanoseconds[7]. As the emission from the probe well can be described by the *Bnp* product, where *B* is the radiative recombination coefficient, *n* and *p* are the electron and hole concentrations, respectively, time-resolved PL reflects the time variation of the electron concentration in the probe well and the signal intensity can be enhanced by increasing the hole concentration in forward bias.

In the time-resolved PL measurement, the excitation wavelength of 405 nm was used similarly to the continuous-wave PL measurement. The streak camera images of time-resolved PL spectra from the *reference* [Fig. 1(c)] and the MQW sample [Fig. 1(b)] using excitation energy of approximately 40 pJ per pulse are shown in Fig. 2(c)-(d). From the figure, a remarkable delay in probe well emission ($\lambda$ = 1060 nm) in the MQW sample, defined as carrier TOF, can be clearly seen to be in nanoseconds at 0.6 V. Without applying forward bias, the probe signal was below the detection limit, indicating that, as mentioned above, TOF measurement without applying bias voltage is not suitable for evaluating quantum-structure solar cells. Note that we intentionally used high excitation power to obtain clear PL images in the high-resolution time-wavelength plots. However, time profiles of PL become broad and the estimation of TOF might be inaccurate when the excitation power is too high (see supplementary Fig. 6)[24]. Therefore for the PL time profiles shown later in Fig. 3-4, the samples with and without MQW were excited with lower energies of 6 pJ and 25 pJ per pulse, respectively. These excitation powers were chosen so that their time profiles are not broadened by the high intensity excitation effect while maintaining the S/N ratio high enough.

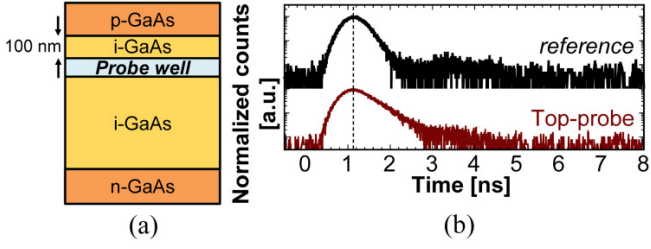

FIG. 3. (a) Structure of the top-probe sample. The structure was similar to the *reference*, but the probe well was inserted at a depth of 100 nm from the top of i-region, shallower than the *reference* by 390 nm. (b) Time-resolved PL from the *reference* and the top-probe sample. No shift in the signal peak indicates negligible transport time in the GaAs bulk region. Although carriers arrive both probes at almost the same time, the discrepancy in the decay shows the different dynamics of carriers after being captured in both probes. This is possibly because of modulation of the electric field in the i-region caused by the background doping concentration of the order of $10^{14}$ cm$^{-3}$. The small peak at 3.4 ns was the artifact from the secondary electrons generated in the photomultiplier tube.

In the *reference*, the quantum structure was replaced by the bulk material with the same thickness $L$. Thus, the measured carrier TOF can be expressed as TOF $= \tau_{\text{Quan}}(L) - \tau_{\text{Bulk}}(L)$, where $\tau_{\text{Quan}}(L)$ and $\tau_{\text{Bulk}}(L)$ are the carrier transport times in the quantum structure and the bulk material with the thickness $L$, respectively. The carrier transport time in the bulk material can be measured by changing the position of the probe well. The result is shown in Fig. 3. There is no shift in the peak of probe signal on moving the probe position by 390 nm, suggesting that the transport time in the GaAs bulk layer is negligible in this case. When the transport time in bulk material is negligible, the approximation TOF $\approx \tau_{\text{Quan}}(L)$ is valid. This implies that the observed delay is a measure of the carrier transport time across the quantum structure contained in the i-region of the solar cell.

In the discussions so far, an assumption was made that the holes that are photoexcited at the p-doped surface are blocked and only the electron transport was observed by the TOF technique. The assumption was verified by examining TOF at various bias voltages because electrons and holes have the opposite behavior. At higher forward bias, the effective potential barrier for electrons and holes in the probe direction became higher and lower, respectively, resulting in slower electron and faster hole transport (see supplementary Fig. 7)[24]. In addition, a forward bias reduces the field that sweeps out electrons and blocks holes distributed at the continuum level. As seen in the results shown in Fig. 4(a), a higher forward bias gives a larger TOF, which is consistent with the electron behavior. Therefore, it is experimentally confirmed that TOF extracted by this technique reflects the electron transport in p-on-n quantum structures.

By defining averaged electron drift velocity $v_{\text{drift}}$ as

$$v_{\text{drift}} = \frac{L}{\text{TOF}}, \quad (1)$$

the dependence of field $E$ on $v_{\text{drift}}$ in the MQW region can be investigated using the results from Fig. 4(a). Note that the MQW thickness is $L = 100$ nm (8 wells and 9 barriers) for this sample. The electric field $E$ across the MQW can be estimated by solving the Poisson and current continuity equations in a bulk GaAs p-i-n junction using the doping concentrations of $1 \times 10^{17}$ and $2 \times 10^{18}$ cm$^{-3}$ in the n-doped base and p-doped emitter, respectively, as first-order approximation. More accurate value of field strength can be obtained by considering carrier dynamics in MQW, which is what we aim to clarify. The plot of the averaged drift velocity against the field is shown in Fig. 4(b). An i-region background doping of the order of $1 \times 10^{14}$ cm$^{-3}$, estimated by the same method in Ref. [21], modulated the field, and the estimated effect is shown as error bars in the figure. The result shows that $v_{\text{drift}}$ increases almost linearly with the field strength, at least until the field strength employed in typical solar cells. The vertical-axis intercept of the fitting line is $(4 \pm 4) \times 10^2$ cm/s, indicating that it tends to pass close to the origin within an error margin. This is an important finding which suggests that in spite of the complicated carrier cascade dynamics inside quantum structures and the field-nonlinearity of carrier escape rate[11], carriers apparently behave as if they are in an equivalent material with a constant *effective drift mobility*

$$\mu_{\text{eff}} = \frac{v_{\text{drift}}}{E}. \quad (2)$$

Based on the slope, the effective electron drift mobility in the MQW is estimated as $\mu_{\text{eff}} = 0.28 \pm 0.04$ cm$^2$/Vs. This method of estimating the effective drift mobility is a significant step forward in modeling and analysis of carrier dynamics in quantum structure devices.

By contrast, the measurement results using the conventional time-resolved PL technique, which focuses on the PL emitted from the MQW itself and has sometimes been used to evaluate carrier transport in quantum structures[13,21-23], at different bias voltages are shown in Fig. 4(c). Despite the variation in the TOF measured via the probe well at forward bias [Fig. 4(a)], the results show similar time-resolved PL emitted from the MQW. In the conventional technique, if the recombination is comparatively fast, the profile of the time-resolved PL emitted from a quantum structure is determined only by the recombination rate and cannot provide information about carrier transport. This shows the superiority of the TOF technique over the conventional one.

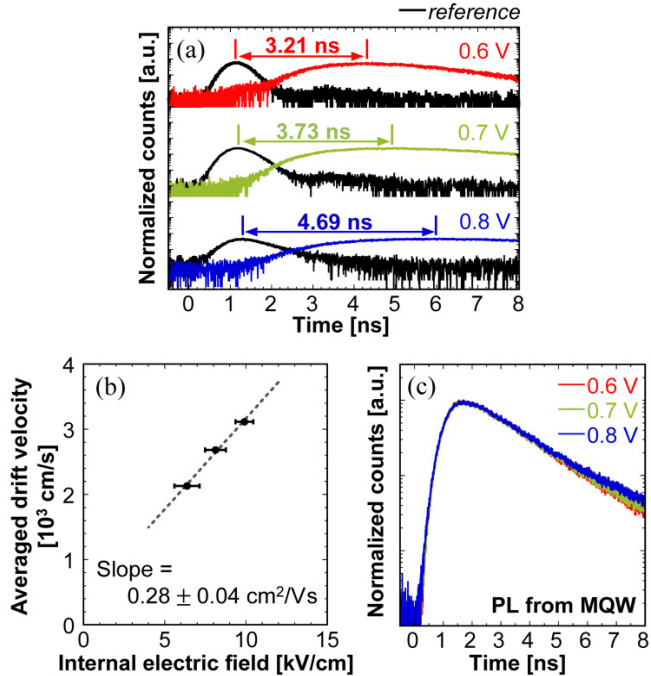

FIG. 4. (a) Time-resolved PL from the probe well at various applied bias voltages. Three slow-delay color profiles represent the probe signals from the MQW sample, while the fast-decay black profiles show the probe signals from the *reference* at the corresponding bias voltages. The peak delays caused by MQW, the TOFs, are indicated in the figure. The standard deviation of the fitted peaks is 0.02 ns. (b) Averaged drift velocity [Eq. (1)] at various electric field strengths. The uncertainties in the electric field values and slope were estimated based on the effect of the unintentionally doped background concentration in the i-region. (c) Time-resolved PL from the MQW structure at various applied bias voltages.

To study the influence of carrier obstruction on device performance, carrier collection efficiencies (CCE) of a GaAs reference and an MQW solar cell were calculated. CCE is a measure of the percentage of carriers collected as photocurrent and is expressed as

$$\text{CCE}(V,\lambda) = \frac{\text{EQE}(V,\lambda)}{\text{EQE}(V_{\text{reverse}},\lambda)}, \quad (3)$$

where $V_{\text{reverse}}$ is the sufficient reverse bias at which almost all of the photoexcited carriers are collected[20]. The structure of the two cells for evaluating CCE were almost the same as that used in the TOF measurement. The only difference was that the probe wells were removed to eliminate their influence. The results are shown in Fig. 5. Under 405-nm-wavelength illumination with intensity of 2.5 mW/cm$^2$, the CCE of the GaAs cell is almost unity, whereas the CCE of the MQW cell reduces by approximately 2% at 0.6 V. This illustrates the correspondence between the microscopic TOF and the macroscopic photocurrent. On the other hand, it is notable that with even for small change in CCE, the carrier TOF is remarkably sensitive to the carrier transport, indicating the suitability of the technique for measuring carrier dynamics in quantum structures.

In summary, we have proposed a carrier TOF technique using a quantum-well probe to observe carrier transport inside quantum-structure solar cells in the microscopic scale. The simplicity for analysis and evaluation accuracy in comparison to the conventional techniques will make the optimization of structure design more straightforward for carrier transport enhancement. Furthermore, this measurement technique and the effective mobility evaluation will help better understanding of carrier dynamics, such as cascade dynamics in superlattice structures, and their effects on the macroscopic current-voltage property, allowing us to maximize the potential of quantum-structure solar cells as well as other quantum-structure device applications. It should be pointed out that even though we have conducted the experiment on MQW, this method can also be applied to quantum dots or other nanostructures.

A part of this study is supported by the Research and Development of Innovative Solar Cell program, New Energy and Industrial Technology Development Organization (NEDO), Japan

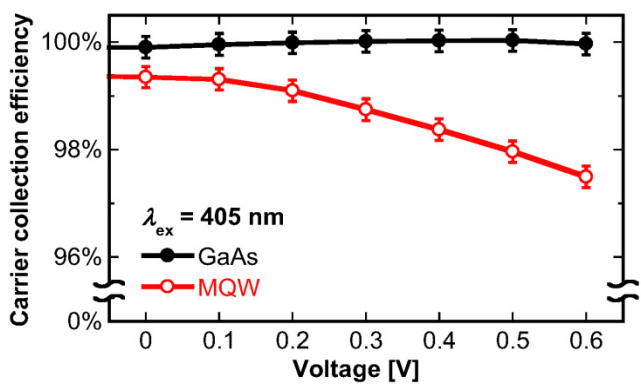

FIG. 5. CCE [Eq. (3)] under 405-nm-wavelength illumination vs. applied bias voltages. The samples used were a GaAs solar cell and an 8-well MQW solar cell, both without a probe well. The error bars were calculated from the fluctuation in the measured photocurrent.